\documentclass[preprint,aps]{revtex4}
\usepackage{graphicx}
\usepackage{mathptmx} 
\usepackage{natbib}

\begin{document}

\newpage

\title{
Universal Complex Structures in Written Language
}
\author
{
\'Alvaro Corral$^1$,  
Ramon Ferrer-i-Cancho$^2$
\&
Albert D\'{\i}az-Guilera$^3$
}
\affiliation{
$^1$%
Centre de Recerca Matem\`atica,
Edifici Cc, Campus Bellaterra,
E-08193 Cerdanyola, Barcelona, Spain.\\
$^2$%
Departament de Llenguatges i Sistemes Informatics,
        Universitat Politecnica de Catalunya,
        Jordi Girona 
        1-3,
        E-08034 Barcelona, Spain.\\
$^3$%
Departament de Fisica Fonamental,
Universitat de Barcelona,
Marti i Franques 1,
E-08028 Barcelona, Spain.
}
\date{\today}



\maketitle
\pagestyle{empty}

{\bf
Quantitative linguistics has provided us with a
number of empirical laws that characterise 
the evolution of languages
\cite{
Lieberman,Atkinson,Wichmann}
and competition amongst them \cite{Wichmann,Abrams}.  
In terms of language \emph{usage}, 
one of the most influential results is Zipf's law of word frequencies
\cite{Zipf1972}.
Zipf's law appears to be universal, and may not even be unique to human language 
\cite{Furusawa2003}.
However, there is ongoing controversy over whether Zipf's law is a 
good indicator of complexity
\cite{Miller1963,Ferrer2006a}.
%
%
Here we present an alternative approach that 
puts Zipf's law in the context of critical phenomena
(the cornerstone of complexity in physics \cite{Stanley_rmp})
and establishes the presence of a large-scale ``attraction'' 
between successive repetitions of words.
Moreover, this phenomenon is scale-invariant and universal -- 
the pattern is independent of word frequency and
is observed in texts by different authors and written in different languages. 
There is evidence, however, that the shape of the scaling relation changes for words that play a key role in the text, implying the existence of different ``universality classes'' in the repetition of words. These behaviours exhibit striking parallels with complex catastrophic phenomena
\cite{Corral_prl.2004,Bunde,Yamasaki}. 
}

Zipf's law states that the relative frequency $f_w$ of any word $w$ in a text 
is approximately related with rank $r_w$ according to an inverse power law 
\cite{Zipf1972}:
\begin{equation}
f_w \propto 1/r_w^\alpha.
\label{Zipf}
\end{equation}
The rank $r_w$ of a word (its position in the list of words ordered by decreasing frequency) measures its rarity. In practice the exponent $\alpha$ is usually close to one. 
Note that Zipf's law describes a \emph{static} property of language; shuffling pages, for instance, would not affect its validity.

To investigate the \emph{dynamical} properties of discourse generation, we can study the distance between consecutive appearances, or ``tokens'', of a given word $w$. This inter-appearance distance, denoted by $\ell$, is measured as the total number of words between two tokens plus one (so it takes on positive values 1, 2, etc.). 
It is immediately clear that $\ell$ can have a very wide range
of values. 
In the novel \emph{Clarissa} by S. Richardson, for example, 
the word \emph{depend} appears 99 times with a mean inter-appearance distance of 9248. 
The minimum distance is 1, and the maximum is 50476.
Although Zipf's law allows one to estimate the mean inter-appearance distance  of each word 
(by means of  $\bar\ell_w \simeq 1/f_w \propto r_w^\alpha$), 
it cannot account for or predict the variability of word recurrence. 
In addition, it has been reported that high-frequency words are distributed 
according to a Poisson process \cite{Herdan1960a}.
This is the simplest stochastic point process, characterised by a total lack of memory. Nevertheless, even the Poisson process is unable to explain such large variability.

In order to characterize the dispersion in $\ell$, we would normally estimate its probability density. Due to the size of the dispersion, however, a large number of appearances are necessary to get significant statistics. But Zipf's law tells us that there are many more low-frequency (high-rank) words than high-frequency (low-rank) words. Thus, the vast majority of words cannot be used in such a study, no matter how long the text considered.

As an alternative, we define a rescaled, ``dimensionless'', 
inter-appearance distance:
$\theta \equiv \ell / \bar \ell_w$. 
For each word $w$, $\theta$
measures the inter-appearance distance in units of its own mean value, 
$\bar \ell_w$. 
We can then calculate the probability density $D_s(\theta)$ 
for large sets of words $s$ having similar relative frequencies $f_w$. 
We remark that we are \emph{not} studying the heterogeneous 
distribution of words in a collection of texts 
\cite{Church,Manning,Montemurro_Zanette}, 
but rather the properties of word repetitions in an individual text. In fact, Zipf himself studied this issue using a similar approach for very low-frequency words \cite{Zipf1972} and proposed a power law for $D_s(\theta)$. As we shall see, the behaviour of $D_s(\theta)$ is actually much more complex.

We have analysed eight texts in four languages 
(for details, see the Methods section). 
The 
curves in Fig. 1 
show the probability densities of rescaled inter-appearance distances 
for all verbs of \emph{Clarissa} appearing in their root form. 
%
%
The verbs are collected into six frequency groups $s$, 
from about 30 appearances to more than 10000,
and $D_s(\theta)$ is plotted for each group in a different point style.
It is clear that the six distributions collapse onto a unique curve. 
This phenomenon signals the approximate fulfilment of a law $D_s(\theta)= F(\theta)$, 
where the \emph{scaling function} $F$ is independent of the word set $s$. It is therefore also independent of frequency, except for the smallest distances ($\ell \lesssim 10$). 
The shape of $F$ approximates a gamma distribution
over about five orders of magnitude,
\begin{equation}
F(\theta) = \frac {1}{a \Gamma(\gamma)} \left( \frac a \theta \right)^{1-\gamma}
e^{-\theta/a},
\label{gamma}
\end{equation}
with the parameter  
$a \simeq 1/\gamma $ (after rescaling, so that $\bar{\theta}=1$)
and $\Gamma$ the Euler gamma function. 
A least-square fit yields $\gamma=0.60 \pm 0.05$.

The validity of this result extends well beyond verbs in \emph{Clarissa}. 
The topmost curves in Fig. 2 
show inter-appearance distance distributions 
for adjectives in the same novel,
well described by the scaling function $F$.
Remarkably, other works in English follow the same trend, 
as shown by the next curves in Fig. 2.
for the adjectives in \emph{Moby Dick} and \emph{Ulysses}. 
Even more unexpectedly, 
the verbs and adjective distributions from texts in French, Spanish, and
Finnish (a highly agglutinative language, in contrast with the other cases) 
display the same quantitative behaviour,
displayed in Fig. 2 and in the Supplementary Information;
in all cases $\gamma$ is in the range $0.60 \pm 0.10$.
It appears that the dimensionless inter-appearance distance distribution 
is ``universal'' in the sense of statistical physics \cite{Stanley_rmp}: 
many different systems (texts) obey the same law despite significant 
differences in their ``microscopic'' details (style, grammatical rules).

Although it has been suggested that Zipf's law may have an elementary 
explanation \cite{Miller1963}, 
clearly the distributions $D_s(\theta)$ are far from trivial. 
This \emph{would} be 
the case if each word followed a Poisson process, as happens in random texts
(see Supplementary Information). 
Instead, the shape of the 
scaling function
implies a clustering phenomenon: 
the appearance of a word tends to ``attract'' more appearances, 
%
%
%
as the distribution is 
dominated by a decreasing power law for $\theta \lesssim 1$. 
This indicates an increased probability for small inter-appearance 
distances relative to the Poisson process,
%
which is characterized by a pure exponential distribution
and approximated by $F(\theta) \simeq 1$ for $\theta < 0.1$. 
The difference is clear between $\ell \simeq 10^{-4} \bar \ell_w$
and $ 0.1 \bar \ell_w$, but beyond $\ell \simeq \bar \ell_w$ 
it is difficult to distinguish the distribution from a Poisson process.

On the other hand, the plot also shows that
clustering and data collapse are not valid for very short
distances, $\ell \lesssim 10$;
rather, some anticlustering (``repulsion'') shows up instead, 
probably due to grammatical and stylistic restrictions on word repetition 
within the same sentence. 
In Figs. 1 and 2, this phenomenon 
is visible as a downturn in many distributions with respect to the 
scaling function on the left-hand side.

Clustering properties of words
have some striking similarities to the occurrence of natural hazards
\cite{Corral_prl.2004,Bunde};
the time delay between earthquakes is shown in the lowest curve of Fig. 2 (``earthQ'')
for comparison. 
This suggests that the models used to describe aftershock triggering may also 
provide insight into the process of text generation:
the appearance of a word enhances its likelihood for a certain time, 
but without a characteristic
scale up to the mean distance $\bar \ell_w$.
This result also validates Skinner's hypothesis regarding the 
repetitive appearance of sounds in speech
\cite{Skinner}.

The case of nouns and pronouns is more intricate.
Their overall distributions $D_s(\theta)$ clearly deviate
from the function $F$.
It turned out that for some words, inter-appearance distances considerably larger 
than the mean value ($\theta \gg 1$) are more common than the scaling relation 
would predict. 
If we remove by hand the relatively few nouns and pronouns with 
anomalous behaviours, 
we recover the same law followed by verbs and adjectives. 
For instance, for \emph{Clarissa} 
one only needs to eliminate 12 nouns and 10 pronouns 
(out of 315 and 34, respectively). 

We now turn our attention to these special words. Surprisingly, they appear to follow a new type of structure. 
We display in Fig. 3a the distributions for 9 nouns 
(\emph{letter, lady, mother, brother, father, sister, uncle,
lord, cousin)} and 6 pronouns (\emph{his, your, her, him, she, he}).
Both
groups are divided into two frequency groups. 
The four distributions still collapse onto a unique curve, 
but the results are clearly not fit by the function $F$. 
Rather, we need a new scaling function $G$. 
A good approximation is the stretched exponential function
\begin{equation}
G(\theta)=\frac { \delta} {a' \Gamma(1/\delta)} \, e^{-(\theta/a')^\delta}.
\end{equation}
Again, rescaling (so that the mean is 1)
fixs one of the two parameters: 
$a'\simeq \Gamma(1/\delta)/\Gamma(2/\delta)$.
The remaining free parameter is $\delta=0.33\pm 0.05$.
The scaling function $G$ also describes the clustering properties of words, but its behaviour is different from that of $F$. Relative to a Poisson process, these words are more likely to occur at both short distances and long distances; the clustering effect is therefore quite pronounced. 
Amazingly, the function $G$ can also be used to describe the times between ups or downs in financial markets \cite{Yamasaki}. 
The remaining 7 nouns and pronouns in \emph{Clarissa}, fit by neither $F$ nor $G$, are 
\emph{colonel, captain, sir, you, I, me,} and \emph{my}
(see Supplementary Information).

We can proceed by considering what these laws mean for individual words. 
If we can accurately measure a number of single-word probability densities 
$D_w(\ell)$, then we can compare their shapes by means of the scale transformation
$\ell \rightarrow \ell/ \bar \ell_w \equiv \theta$
and
$D_w(\ell)  \rightarrow \bar \ell_w D_w(\ell)$. 
A sufficient condition for the 
collapse of $D_s(\theta)$ to hold is that
\begin{equation}
D_w(\ell) = G(\ell/\bar \ell_w)/\bar \ell_w
\label{scaling}
\end{equation}
(the same holds for $F$).
This scaling law is shown to be very accurate in Fig. 3b, 
which displays the individual distributions for the nouns and pronouns 
whose averaged distributions were shown in Fig. 3a. The extension of this law to other texts and languages is demonstrated in Fig. 3c, with impressive results.

If we now relate the mean distance to Zipf's law (\ref{Zipf}),
$\bar\ell_w \simeq 1/{f_w} \propto r_w^\alpha$,
then Eq. (\ref{scaling}) becomes
\begin{equation}
D_w(\ell) =  \tilde G(\ell/r_w^\alpha)/r_w^\alpha
\label{scaling2}
\end{equation}
where $\tilde G$ is essentially $G$
after including the normalization constant of Zipf's law.
This is just the condition of scale invariance for
functions with two variables \cite{Christensen_Moloney},
and reflects the signature characteristic of word repetition: 
each word follows the same pattern, although on a different scale which depends on its average frequency.
In other words, different parts of a text (word occurrences) have the same structure 
despite great differences in their scales (this is also the main characteristic of fractals). 
In statistical physics, distinct scaling relations such as $F$ and $G$ (or distinct values of the exponent $\alpha$) define \emph{universality classes} \cite{Stanley_rmp}. In linguistics these universality classes comprise different \emph{types} of words, independent of author and language.

Going back to the case of adjectives and verbs, 
it turns out that the single-word distributions are described by $F$ 
only on average; 
that is, in many cases there are deviations between their rescaled  
$D_w(\ell)$ and $F(\theta)$. 
The small deviations appearing at large $\theta$ in Fig. 2 may originate
from special cases, for instance the word \emph{belle} (beautiful) in \emph{Artam\`{e}ne}. The distribution of this word actually scales nicely with $G$, as seen in Fig. 3c. 
Nevertheless, such cases are rare and their statistical weight is so low that they barely modify the shape of $F$.

We may wonder if the universality classes that describe word clustering are 
an intrinsic property of language itself, or rather a fundamental characteristic 
of human behaviour reflected in literary works \cite{Barabasi,Harder_Paczuski}. 
The first possibility is favoured by the fact that adverbs and even function words 
(conjunctions, prepositions, and determiners) clearly do not follow a Poisson process, 
except perhaps for those words with the very lowest ranks 
(see Supplementary Information).
Higher-rank adverbs and function words display clustering, 
and are well described by the scaling function $F$.

However, we can establish at least one important distinction between 
the universality classes we have found. 
It turns out that most of the ``special'' nouns distributed according to $G$ refer to persons with a particularly relevant role
(note that nouns not referring to persons 
can also play an important role, as is the case 
with \emph{letter} in \emph{Clarissa}, an epistolic novel). 
Something similar is clearly happening with pronouns, as the special cases are always personal or possessive. 
We observe increased clustering for these key words, 
although unlike Ref. \cite{Ortuno2002a} 
we find well-defined universality classes. 
This supports the idea that this kind of clustering originates in the special properties of human behaviour \cite{Barabasi,Harder_Paczuski}.

{\bf METHODS SUMMARY}


{\bf Identification of parts of speech.}
English words were placed into grammatical categories using A. Kilgarriff's lemmatised list, elaborated from the British National Corpus \cite{Kilgarriff}. However, we have changed the classification of possessive determiners to possessive pronouns \cite{Manning} as their behaviour is consistent with the clustering properties obtained for other pronouns. Words belonging to more than one category were excluded from the study. 
For Spanish words, we mainly used 
the Wictionary from Wikipedia \cite{Wiki}
but also drew on 
the electronic dictionary built by L. Padr\'{o} \emph{et al}. 
for FreeLing \cite{Padro}.
For French, we made use of the list elaborated by S. Sharoff \cite{Sharoff}, 
and Finnish words were identified from a list available at the CSC, the Finnish
IT Centre for Science \cite{Finnish}.

\bibliographystyle{nature}


{\bf Acknowledgements}
We are indebted to L. L\"{o}fberg and L. Padr\'{o} for assistance
and to G. Altmann
and H. Rodr\'{\i}guez
for their comments on the manuscript.
The authors participate in different research projects
funded by Spanish and Catalan agencies.

{\bf Author Information}
 Correspondence and requests for materials should be
addressed to A.C. (ACorral@crm.cat).

\newpage

{\bf METHODS}

{\bf Titles analyzed.}
The following table provides details on the texts. 
The labels (column 1) are used in Fig. 2 to identify the curves. 
The date given is the year of first publication, 
and the length of texts is in millions of words (Mw). 
Electronic versions of the texts were downloaded 
from the Gutenberg Project's web page 
\cite{Gutenberg}, except for \emph{Artam\`ene}
\cite{Artamene}.

\begin{tabular}{llllrr}
\hline
label & title & author & language & year & length (Mw) \\
\hline
Clar & Clarissa & S. Richardson & English & 1748 & $0.976  $  \\
Moby & Moby Dick & H. Melville   & English & 1851  & $ 0.215  $  \\
Uly  & Ulysses    & J. Joyce   & English & 1918 & $  0.269  $  \\
Arta & Artam\`ene & Scud\'ery siblings & French & 1649 & $  2.088  $  \\
Brag & Le Vicomte & A. Dumas  & French & 1847  & $ 0.699  $  \\
&      de Bragelonne &  &  &     \\
DonQ & Don Quijote & M. Cervantes & Spanish & 1605   & $  0.381  $  \\
LaRe & La Regenta & L. A. ``Clar\'{\i}n'' & Spanish & 1884 & $  0.308  $  \\
Keva&  Kev\"{a} ja & J. Aho & Finnish & 1906& $0.114  $ \\
& Takatalvi & & & & \\
\hline
\end{tabular}

\bibliographystyle{nature}

\newpage

\begin{figure*}
\caption{
{\bf Scaling and clustering of
inter-appearance distance distributions.}
Verbs appearing in their root form in \emph{Clarissa} 
are considered.
Each distribution includes all words falling 
into one of six equal logarithmic ranges of absolute frequency $n_w$. 
All the distributions are well described by a unique shape: 
the gamma distribution $F$ explained in the text with 
$\gamma=0.60$ and $a=1/\gamma$ (solid line). 
The exponential function, characteristic of Poisson processes,
is shown for comparison.
}
\end{figure*}

\begin{figure*}
\caption{
{\bf Universality of
inter-appearance distance distributions.}
The adjectives in several novels in different languages are analysed 
(see Methods),
except for Finnish (``Keva''), where we have considered verbs,
due to their better statistics. 
From top to bottom the distributions are multiplied by 
$1, 10^{-2}, 10^{-4}$, and so on, 
to avoid overlapping the curves. 
Recurrence-time distributions for earthquakes (earthQ) 
in Southern California are included at the bottom for comparison 
(the distributions include all events with magnitude 
$M \ge M_c$ between 1995 and 1998, where   
$M_c=$ 2, 2.5, 3, and 3.5 \cite{Corral_prl.2004}).
}
\end{figure*}

\begin{figure*}
\caption{
{\bf 
Scaling in a different universality class
for special nouns and pronouns. 
}
(a) Average rescaled probability densities for the 9 nouns and 6 pronouns 
from \emph{Clarissa} listed in the text, where each group is divided into two frequency ranges. 
A stretched exponential function $G$ with $\delta=0.33$ describes all four distributions well. 
The scaling function $F$ is also displayed for comparison. 
(b) Corresponding rescaled distance distributions for individual words.
(c) Rescaled distance distributions for words from other works, including the adjective \emph{belle}.
}
\end{figure*}

\newpage

\begin{figure*}
\centering
\includegraphics[height=8cm]{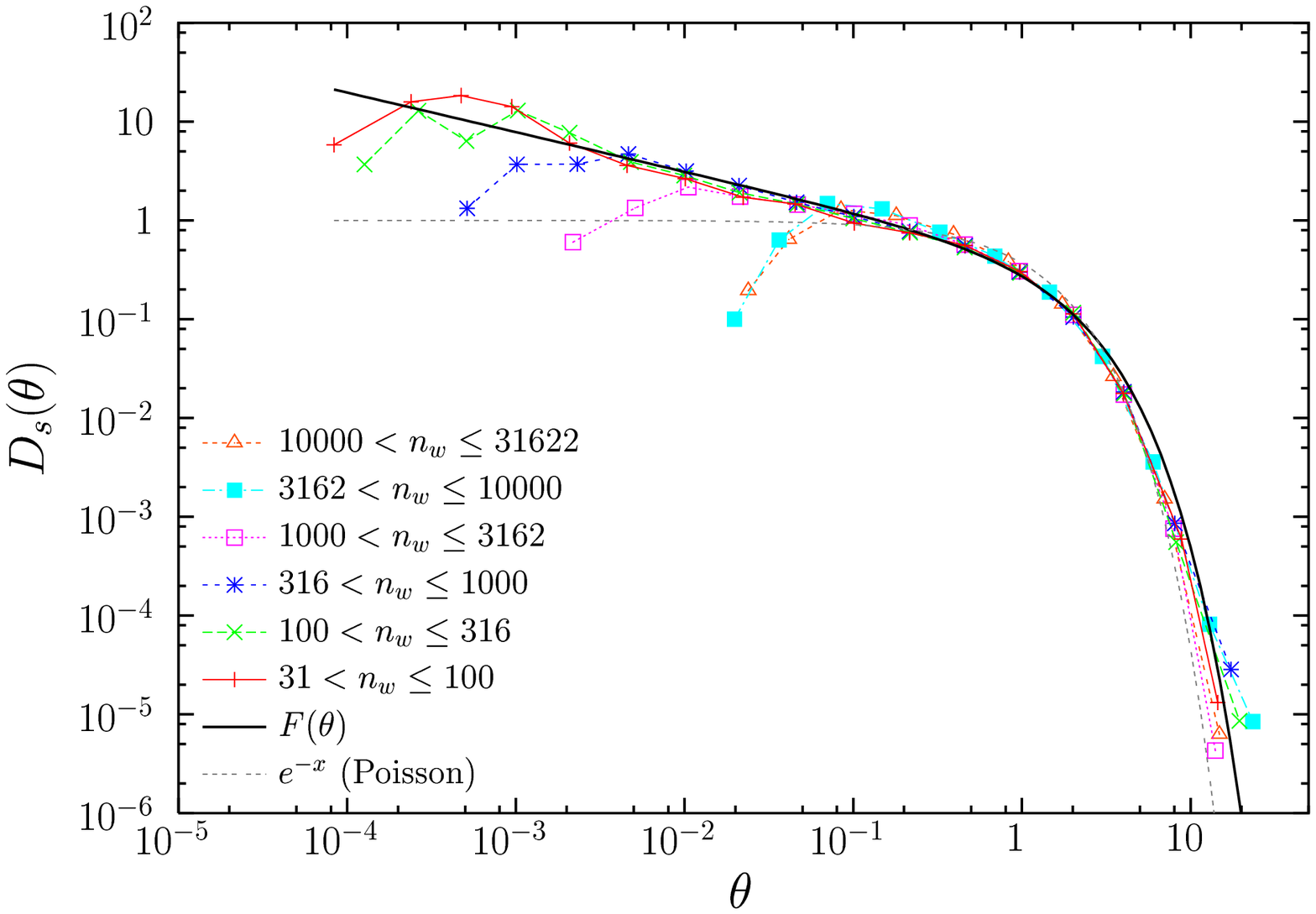}\\
FIG. 1
\end{figure*}

\begin{figure*}
\centering
\includegraphics[height=20cm]{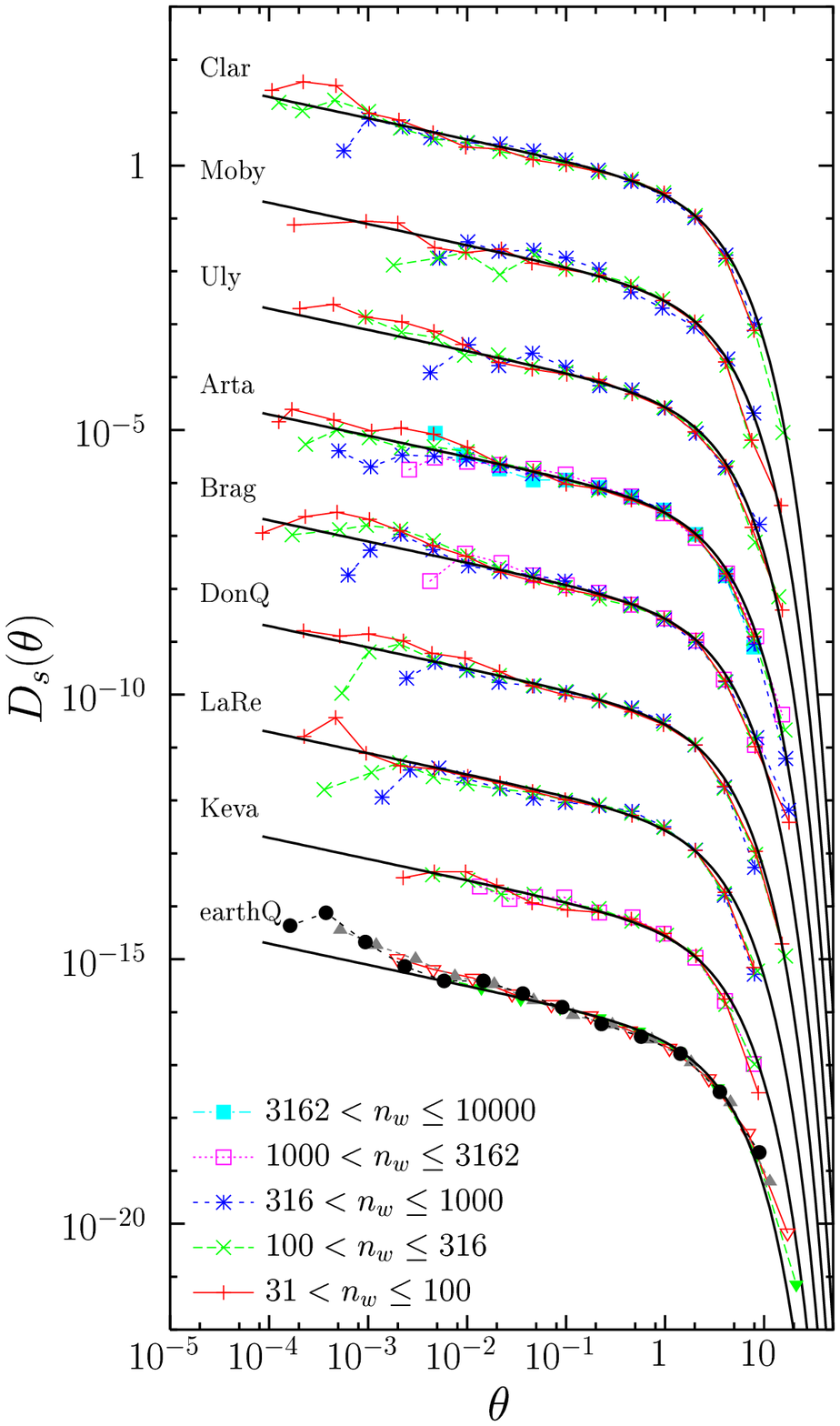}\\
FIG. 2
\end{figure*}

\newpage
\begin{figure*}
\centering
{\Large \hspace{-8cm} a}\\
\includegraphics[height=7.5cm]{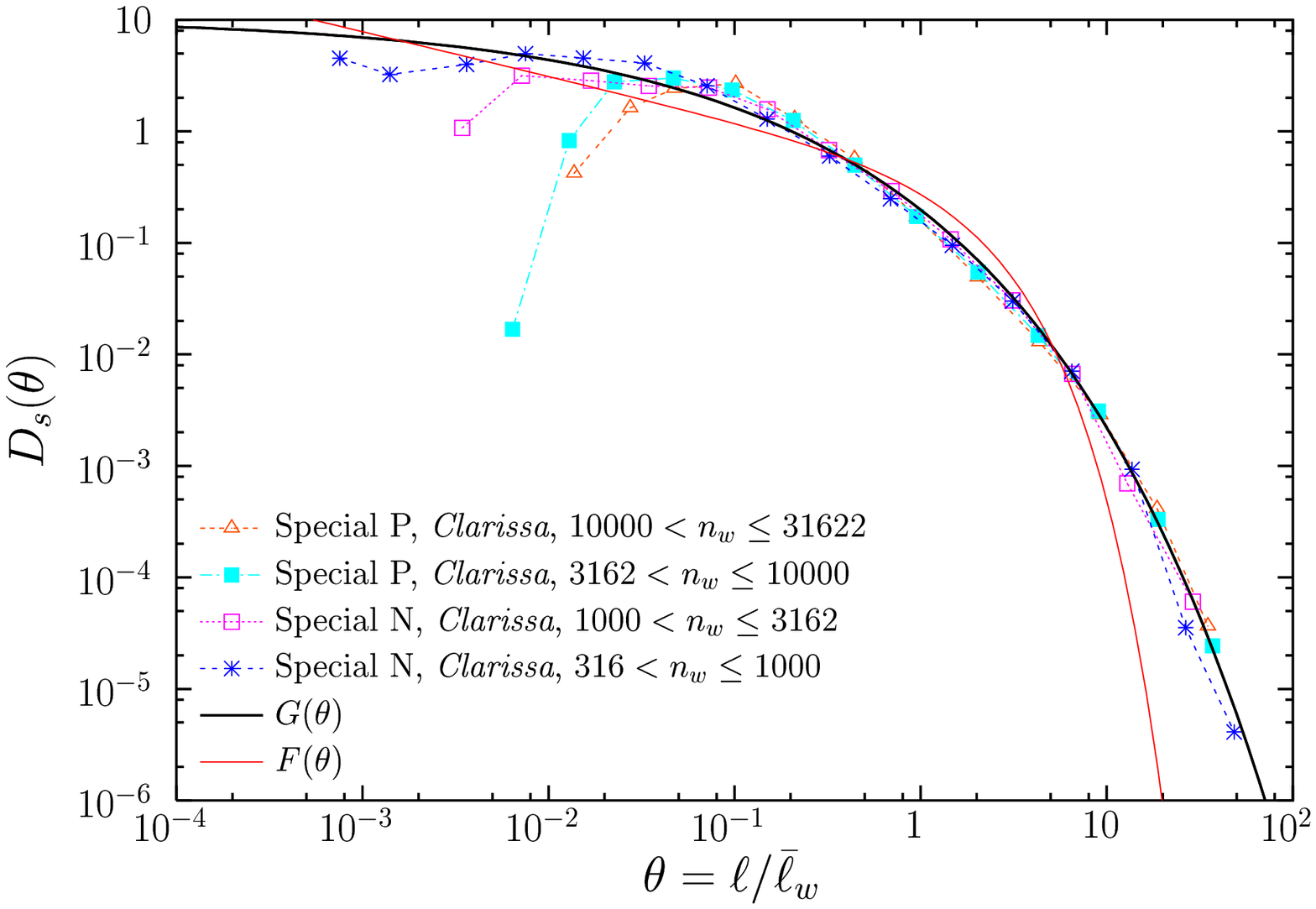}\\
{\Large \hspace{-8cm} b}\\
\includegraphics[height=7.5cm]{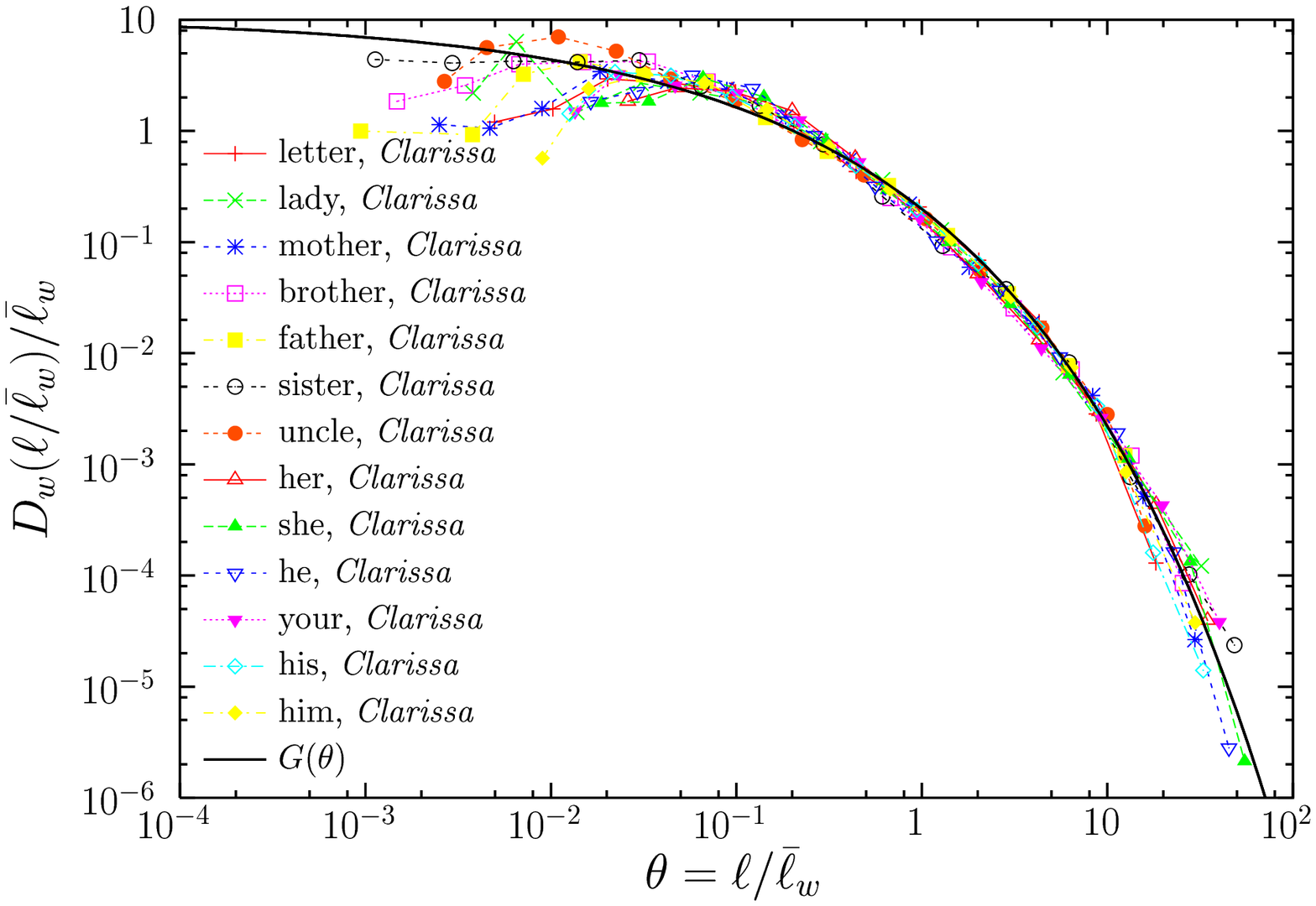}\\
{\Large \hspace{-8cm} c}\\
\includegraphics[height=7.5cm]{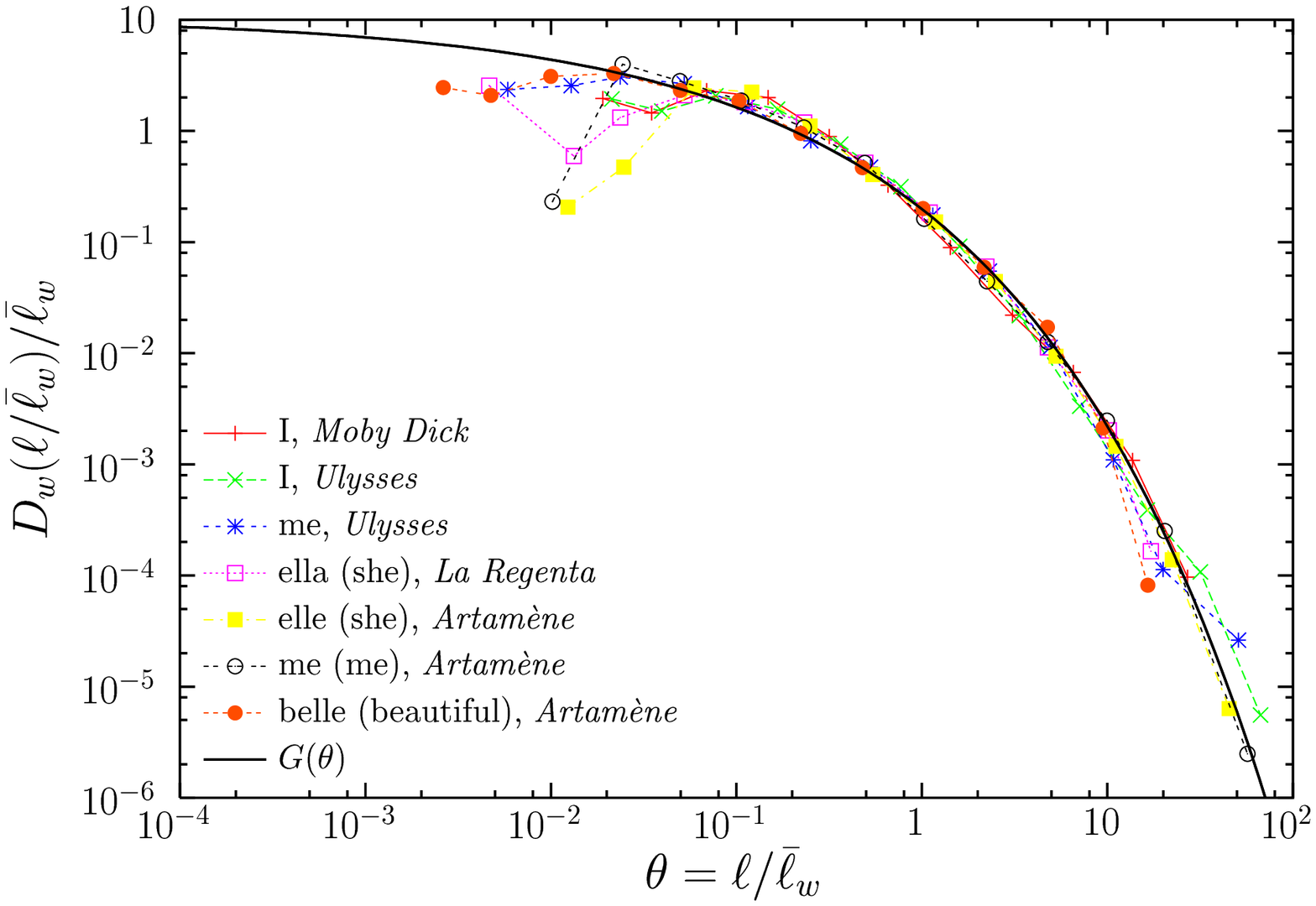}\\
FIG. 3
\end{figure*}

\end{document}